\documentclass[a4,aps,pre,twocolumn,10pt]{revtex4-1}
\usepackage{amsmath}
\usepackage{amsfonts}
\usepackage{amssymb,enumerate}
\usepackage{graphicx,hyperref}
\usepackage{amssymb}
\usepackage{epstopdf,booktabs}
\usepackage{mathptmx}
\usepackage[usenames,dvipsnames]{color}
\usepackage[T1]{fontenc}
\usepackage[utf8]{inputenc}
\usepackage{natbib}
\newcommand{\ptd}[1]{\frac{\partial #1}{\partial t}}
\newcommand{\ttd}[1]{\frac{d #1}{dt}}
\newcommand{\ptdd}[1]{\frac{\partial^2 #1}{\partial t^2}}
\newcommand{\psd}[1]{\frac{\partial #1}{\partial z}}
\newcommand{\psdinline}[1]{\partial #1/\partial z}
\newcommand{\psdd}[1]{\frac{\partial^2 #1}{\partial z^2}}
\newcommand{\pstd}[1]{\frac{\partial^2 #1}{\partial t\partial z}}

\newcommand{\be}{\begin {equation}}
\newcommand{\ee}{\end {equation}}
\newcommand{\beqa}{\begin {eqnarray}}
\newcommand{\eeqa}{\end {eqnarray}}
\newcommand{\mb}{\mathbf}
\hypersetup{
    colorlinks,    citecolor=blue,    filecolor=black,    linkcolor=blue,    urlcolor=black
}

\begin{document}
\title{Ion acceleration `via' relativistic self induced transparency in subwavelength target}
\author{Shivani Choudhary}
\email{shivani.choudhary@pilani.bits-pilani.ac.in}
\affiliation{Department of Physics, Birla Institute of Technology and Science - Pilani, Rajasthan, 333031, India}
\author{Amol R. Holkundkar}
\email{amol.holkundkar@pilani.bits-pilani.ac.in}
\affiliation{Department of Physics, Birla Institute of Technology and Science - Pilani, Rajasthan, 333031, India}
 
\begin{abstract}
In this work we have studied the effect of target thickness on relativistic self-induced transparency (RSIT) and found out that for subwavelength targets the corresponding threshold target density (beyond which target is opaque to incident laser pulse of a given intensity) increases. The accelerating longitudinal electrostatic field created by RSIT from subwavelength target is used to accelerate the ion bunch from a thin, low density layer behind the main target to $\sim$ 100 MeV. A suitable scaling law for optimum laser and target conditions  is also deduced. The word `via' in the title signifies the fact that we are interested in acceleration of ions from the layer placed behind the target. It is also being observed that as per as energy spectrum is concerned; an extra low density layer is advantageous than relying on target ions alone.  
 
\end{abstract}

\maketitle

\section{Introduction}
 
The laser based particle acceleration have been shown a great deal of research interest  since last couple of decades and still continues to do so. The researchers in this field around the globe are working towards the understanding and the development of the efficient acceleration mechanism both on experimental and theoretical front. The development of efficient particle beam source promises remarkable applications in various areas of applied sciences \cite{cancer,Cancer2013174,ldp,hp,PhysRevLett.86.436,proton}. 

The laser and target conditions play paramount role in deciding which acceleration mechanism would prevail. Among all the
acceleration mechanisms the Target Normal Sheath Acceleration (TNSA) is studied extensively both on experimental and theoretical
front \cite{PhysRevLett.85.2945,:/content,nature,PhysRevE.85.036405,tnsa,wilks2001_pop,kim2016_pop}. The weak scaling of proton energy with laser intensity ($\sim \sqrt{I_0}$) is a severe limitation of the TNSA as per as GeV protons are concerned \cite{PhysRevLett.85.2945}. When the target thickness is comparable to skin depth of the plasma, then after TNSA another mechanism namely Break-Out Afterburner (BOA) takes over. In this stage, the hot electron plasma expands sufficiently by laser pondermotive force so that target becomes underdense. The laser penetrates the target and strong electric field is induced through Buneman instability, this growth of instability results in conversion of electron energy to ion energy and thus, ions are accelerated to very high energies \cite{BOAultrathin,LPB444856,yin2011_pop}.  

When the laser intensity increases to $\sim 10^{20}$ W/cm$^2$ then radiation pressure acceleration (RPA) starts to dominate the acceleration process  \cite{bqmz,hb,PhysRevLett.109.185006,PhysRevSTAB.12.021301,chen_csa,PhysRevLett.109.215001,PhysRevLett.92.015002,kar_rpa}. Depending on the target thickness two mechanisms can occur under RPA viz, Hole Boring (HB) for thick targets and Light Sail (LS) for thin targets. The RPA is mainly governed by circularly polarized laser pulse so that $\mb{j}\times\mb{B}$ heating can be avoided and electrons are pushed deep into the targets. In HB radiation pressure drives material ahead of it as a piston but does not interact with the rear surface of the target \cite{0741-3335-51-2-024004,0741-3335-54-11-115001}. However, in alternate scenario if the target is sufficiently thin for the laser pulse to punch through the target and accelerate a slab of plasma as a single object is referred as LS regime \cite{macchi_ls,macchi_njp}. A wonderful review on this contemporary field of laser induced ion acceleration is presented by Macchi et. al. \cite{RevModPhys.85.751} in which all the ion acceleration mechanisms are covered in depth. 

In order to efficiently accelerate the ions to ultra-relativistic energies the use of multilayer or multi-species target is also examined
by various researchers \cite{1367-2630-11-9-093035,PhysRevLett.105.065002,PhysRevSTAB.12.021301,:/content/aip/journal/pop/16/8/10.1063/1.3196845,:/content/aip/journal/pop/18/5/10.1063/1.3574388}. In Ref. \cite{PhysRevE.78.026412} authors have shown the acceleration of protons to $\sim 210$ MeV using double layered thin foil by the laser with peak intensity $2.7\times 10^{22}$ W/cm$^2$. In this so called Directed Coulomb Explosion (DCE), the high intense laser
beam interacts with very high density Al$^{13+}$ plasma ($\sim 400\ n_c$) which expels all the electrons from the target leaving behind 
ions, which undergoes Coulomb explosion, the adjacent proton layer of $30n_c$ then gets accelerated by the electrostatic fields created 
by the DCE \cite{PhysRevSTAB.12.021301}.

Each mechanism defers from another on the basis of laser and target conditions used. The relativistic self induced transparency (RSIT) can also be perceived as one of the mechanism for acceleration of ions, this exploits the fact that the threshold plasma density for laser penetration, increases as a consequence of relativistic mass effect of the plasma electrons. RSIT was initially reported in pioneer work of Refs. \cite{kaw,PhysRevLett.27.1342}, since then it has also drawn considerable interest in ion acceleration fraternity. RSIT can prevent the RPA in thin targets, and hole-boring in thick targets. Furthermore, RSIT can results in hot electrons and thus allowing for more energetic ion acceleration \cite{:/content/aip/journal/pop/17/4/10.1063/1.3368791, PhysRevLett.102.125002,LPB:10039071}. Recently laser-driven Neutron source based on the relativistic transparency of solids is also being reported \cite{PhysRevLett.110.044802}.
 
As per as RSIT is concerned, many of the theoretical discussions seek the steady state solution in the purview of relativistic cold fluid model. However, in this work we attempted to solve the set of dynamical equations in space and time which governs the response of semi-infinite plasma to intense laser fields. In order to study the RSIT for thin targets, we relied on relativistic stationary plasma model along with PIC simulations. In this work we would be using RSIT phenomenon to accelerate the protons to very high energies using a secondary thin and low density layer behind the main target. The protons from this secondary layer will then be  accelerated as a bunch in the electrostatic field created by RSIT mechanism in main target. The dependence on the target thickness and density is also being studied and it has been observed that the threshold density for RSIT increases for subwavelength targets. The objective of this article is to understand the formation of electrostatic fields by RSIT mechanism and its usefulness to accelerate protons. In section II we briefly discuss the RSIT mechanism along with a theoretical and simulation model for the same. The detailed formulation of the dynamical equations are discussed in Appendix \ref{App-A}. Results and discussions are presented in section III followed by the concluding remarks in section IV.

\begin{figure*}
\centerline{\hspace{0.2cm}\includegraphics[scale=0.9]{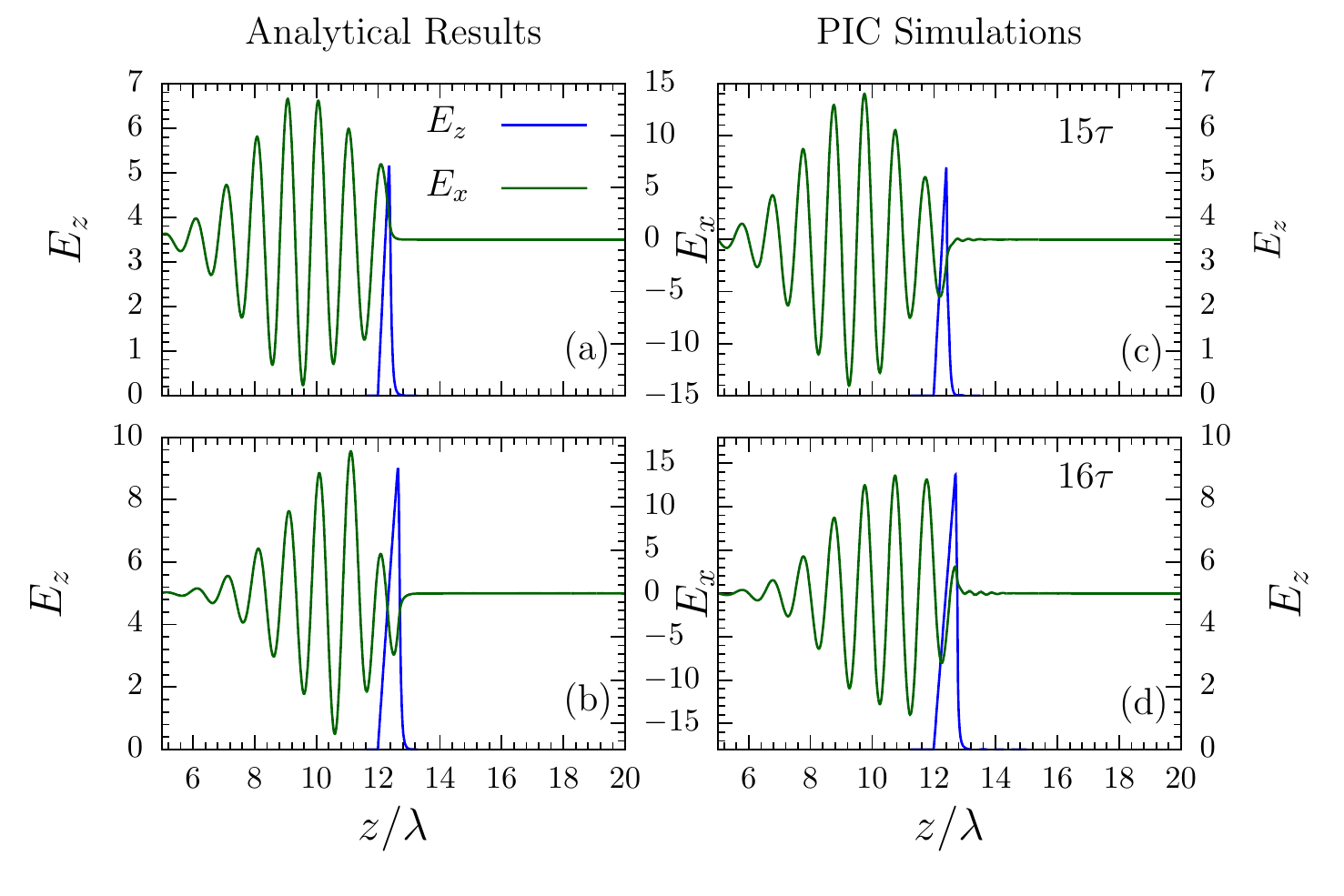}}
\caption{(color online) The component of the laser field along $x$ direction ($E_x$) and electrostatic field ($E_z$) created by charge separation is presented by numerically solving Eqs. \ref{phi0} - \ref{ga0} (left panel) and compared with 1D PIC simulations (right panel). }
\label{rsit_ana}
\end{figure*} 
 
\section{Theory and Simulation Model}
The propagation of electromagnetic (EM) waves in plasmas have been studied quite extensively and vast literature on the topic can be found \cite{bittencourt,chen}. The dispersion relation for EM waves propagation in plasmas is expressed as 
\be \omega^2 = \omega_p^2 + k^2c^2, \label{dis}\ee 
where $\omega$, and $k$ are respectively the frequency and wave propagation vector of EM waves, $c$ is speed of light in vacuum and $\omega_p = \sqrt{n_e e^2/\varepsilon_0m_e}$ is the natural frequency of the plasma oscillations. Here $e$, $m_e$ and $n_e$ are respectively the electron charge, electron mass and electron density. 

It can be understood from Eq. \ref{dis} that EM waves can not propagate beyond the point where $\omega_p > \omega$ because it meant the wave vector to be imaginary, which is physically equivalent of attenuation of EM waves. In terms of the plasma density it means that EM wave can not propagate in plasma beyond a critical density $n_c$ which is defined as $n_c = \varepsilon_0 m_e \omega^2/e^2$ (it is the density for which the plasma frequency matches to that of EM wave), however for densities $n_e < n_c$, the EM wave can propagate without much of attenuation. 

The interaction of very intense laser fields can change the above mentioned criteria about the critical density. In such situations the relativistic mass of the electrons need to be taken into account, which in principle increases the threshold density ($n_c$) for the propagation of laser for a given intensity. This effect is refereed as relativistic self induced transparency (RSIT). The interaction of ultra intense laser pulses with plasma can be modeled by solving the wave propagation 
equation in Couloumb gauge, continuity equations, posisson's equation along with relativistic Lorentz force
equations. The analytical treatment of RSIT based on relativistic stationary plasma model (cold fluid theory with steady-state solutions) is being vastly discussed in the litretaure \cite{PhysRevE.62.1234,LPB:67593, PhysRevLett.87.275002}. However, in this work we have attempted to solve the set of dynamical equations in space and time which governs the plasma response to intense laser fields. We would restrict ourselves to the one dimensional scenario where laser is considered to be propagating along $z$ direction, in view of this the spatial derivatives along $z$ directions is only considered. The detailed derivation describing the plasma response (electron + ions) to intense laser fields is discussed in Appendix \ref{App-A}. In the absence of the ionic motion the closed set of equations describing
the plasma response reduces to (Appendix \ref{App-A}),
\be\label{phi0} \psd{\varphi} = - E_z\ee
\be \psd{E_z} = Z n_i - n_e \ee
\be \ptd{E_z} = n_e \frac{p_z^e}{\gamma_e}  \ee
\be \label{wave0}\psdd{\mb{a}} -  \ptdd{\mb{a}} =  \frac{n_e}{\gamma_e} \mb{a}\ee
\be\label{elemotion} \ptd{p_z^e} = - E_z - \psd{\gamma_e} \ee
\be \ptd{n_e} + \frac{\partial}{\partial z}\Big(n_e \frac{p_z^e}{\gamma_e}\Big) = 0 \ee
\be\label{ga0} \gamma_e  = \sqrt{1 + \mb{a}^2 + (p_z^e)^2} \ee
where, $Z$ is atomic number of the ionic species, $n_e$ and $n_i$ are electron and ion densities in units of 
critical density, $p_z^e$ is longitudinal electron
momentum, $E_z$ is longitudinal electrostatic field created by charge separation, $\gamma_e$ is relativistic
factor of electron, and $\mb{a}$ is vector potential associated with laser pulse. It should be noted that the 
motion of the electrons (Eq. \ref{elemotion}) are governed by the electrostatic field and the ponderomotive 
force ($\psdinline{\gamma_e}$) of the laser pulse (see Appendix \ref{App-A}), which in turn manifests the lowering of the target density and hence assisting the laser propagation deeper into the target, as can be inferred from Eq. \ref{wave0}. Furthermore, it can also be seen from Eq. \ref{wave0} that in if $n_e/\gamma_e \rightarrow 0$ then it will be having a solution of EM wave 
propagation in vacuum. In order to have a RSIT regime the condition should be $n_e/\gamma_e < 1$ or the laser would be reflected from the overdense plasma.  As can be seen from Eq. \ref{ga0} that one in principle needs to solve for $p_z^e$ to
have the knowledge of the $\gamma_e$ and hence the threshold plasma density where RSIT ceases to exist. In this context
one can infer the fact the electron heating ($p_z^e$) plays quite a crucial role in RSIT. 
  
In order to study the interaction dynamics in RSIT regime, we have solved these equations numerically in space and time with following initial/boundary conditions, 
\be \mb{a}(0,t) = a_0 \exp\Big(\frac{-4\ t^2}{T^2}\Big) \Big[ \delta \cos(t) \mb{e_x} + \sqrt{1 - \delta^2} \sin(t) \mb{e_y}  \Big] \ee
where, $T$ is FWHM pulse duration, and $\delta$ is a constant which is either 1 for linearly polarized (LP) laser pulse or 
$1/\sqrt{2}$ for circularly polarized (CP) laser pulses. In this study we will be using only CP laser pulses. In order to 
benchmark the results of this analytical framework, we have simulated the interaction of 5 cycle 
Gaussian laser pulse ($a_0 = 20$) with semi-infinite plasma. The laser is allowed to incident normally from vacuum ($z < 12\lambda$) to a semi-infinite plasma slab ($z\geq 12\lambda$) with density $n_i = n_e = 2 n_c$. 

In Fig. \ref{rsit_ana} we have compared the longitudinal electrostatic field ($E_z$) as calculated by numerically solving Eq. \ref{phi0} to \ref{ga0} with 1D PIC simulations, and as can be seen from the same that the outcome of theoretical formulation is reasonably in good agreement with PIC simulations. However, in order to study the laser interaction with target of finite thickness, one needs to incorporate the plasma expansion in vacuum \cite{PhysRevLett.90.185002} with above set of equations, which is currently beyond the scope of the present work, and we would like to reserve the same for the future.  

Nevertheless, in order to have some qualitative understanding of the effect of finite target thickness and corresponding 
threshold density (beyond which the target will not be transparent for a given laser pulse profile), one can in-principle
seek steady state solution in the transparency regime (other case might be the reflection of incident laser pulse
by overdense plasma). Except where otherwise stated, we will be studying the interaction of flat top laser pulse ($a_0$) of duration 6 cycles with rise and fall of 1 cycle each, with a target of thickness $d$ and density $n_e$. The pulse profile is considered to be flat top, such that the condition for RSIT ($n_e/\gamma_e < 1$) can be met from the instant laser hits the target. Recently in Ref. \cite{PhysRevE.86.056404}, the relativistic cold fluid model with stationary solutions have been used for semi-infinite plasma and an expression for threshold plasma density has been reported.  The threshold density for RSIT for high laser intensities is then given by \cite{PhysRevE.86.056404},
\be n_{th} \simeq \frac{2}{9}  \Big[3 + \sqrt{9\sqrt{6}a_0 -12} \Big] n_c \label{nc1}.\ee   
We are mainly interested to have the relativistic transparency in a target with some finite thickness, such that the 
laser should  pass through the target without much of attenuation.

\begin{figure}[t]
\centerline{\hspace{0.2cm}\includegraphics[scale=0.6]{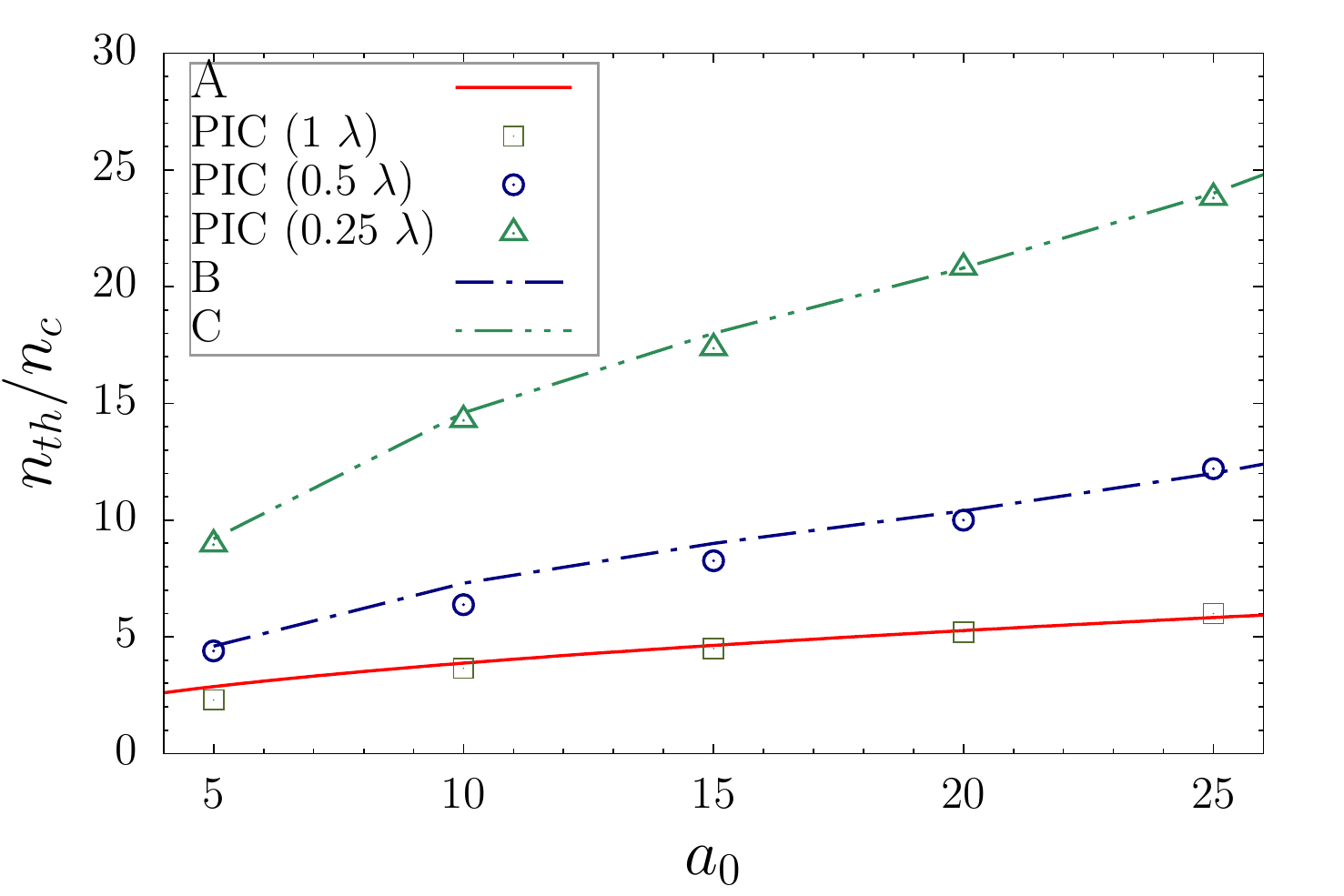}}
\caption{(color online)  Variation of threshold electron density ($n_{th}$) with laser amplitude is presented. Curve `A' shows the expression given by Eq. \ref{nc1}, points denote the results obtained by the 1D fully relativistic PIC simulations for a target thickness of 1$\lambda$ (square), 0.5$\lambda$ (circle) and 0.25$\lambda$ (triangle). Curves `B' and `C' respectively show the qualitative estimate of threshold density for $d = 0.5\lambda$ and $d = 0.25\lambda$ by using Eq. \ref{nc2}.}
\label{threshold}
\end{figure}

Though, the dependence of the target thickness is not explicitly appearing in Eq. \ref{nc1}, but we have tested this expression against a target thickness of 1$\lambda$ using 1D PIC simulation and results are presented in Fig. \ref{threshold}. It can be seen from Fig. \ref{threshold}, that the threshold plasma density for a target of thickness 1$\lambda$ is in good agreement with Eq. \ref{nc1}, which in a sense indicates that the laser of amplitude $a_0$ would be sufficient enough to sweep out all the electrons from a target of thickness 1$\lambda$ and having maximum density of $n_{th}$ as predicted by Eq. \ref{nc1}. Total number of electrons present in a plasma slab (1D scenario) of density $n_e$ and of thickness $d$ should be $n_e d$, so the agreement of Eq. \ref{nc1} and PIC simulation will dictate that laser of amplitude $a_0$ would be sufficient to sweep out
$n_e d$ electrons. We can extend this idea to subwavelength targets, and alternatively can write Eq. \ref{nc1} such that for $d = 1\lambda$ the threshold density is given by Eq. \ref{nc1}. The corrected expression for threshold density for a target
of thickness $d$ should read as,
\be n_{th} \simeq \frac{2 \lambda}{9 d}  \Big[3 + \sqrt{9\sqrt{6}a_0 -12} \Big] n_c \label{nc2}.\ee   
The results obtained by this simple scaling law given by Eq. \ref{nc2} is also presented in Fig. \ref{threshold}, and
compared with PIC simulations for the same. As can be seen that the threshold density as predicted by Eq. \ref{nc2}
is reasonably in good agreement with PIC simulations. It should be noted that, in Fig. \ref{threshold} the region below the data points for particular target thickness would be transparent and the region above the same would be opaque.  

\begin{figure}[b]
\centerline{\hspace{0.2cm}\includegraphics[scale=0.6]{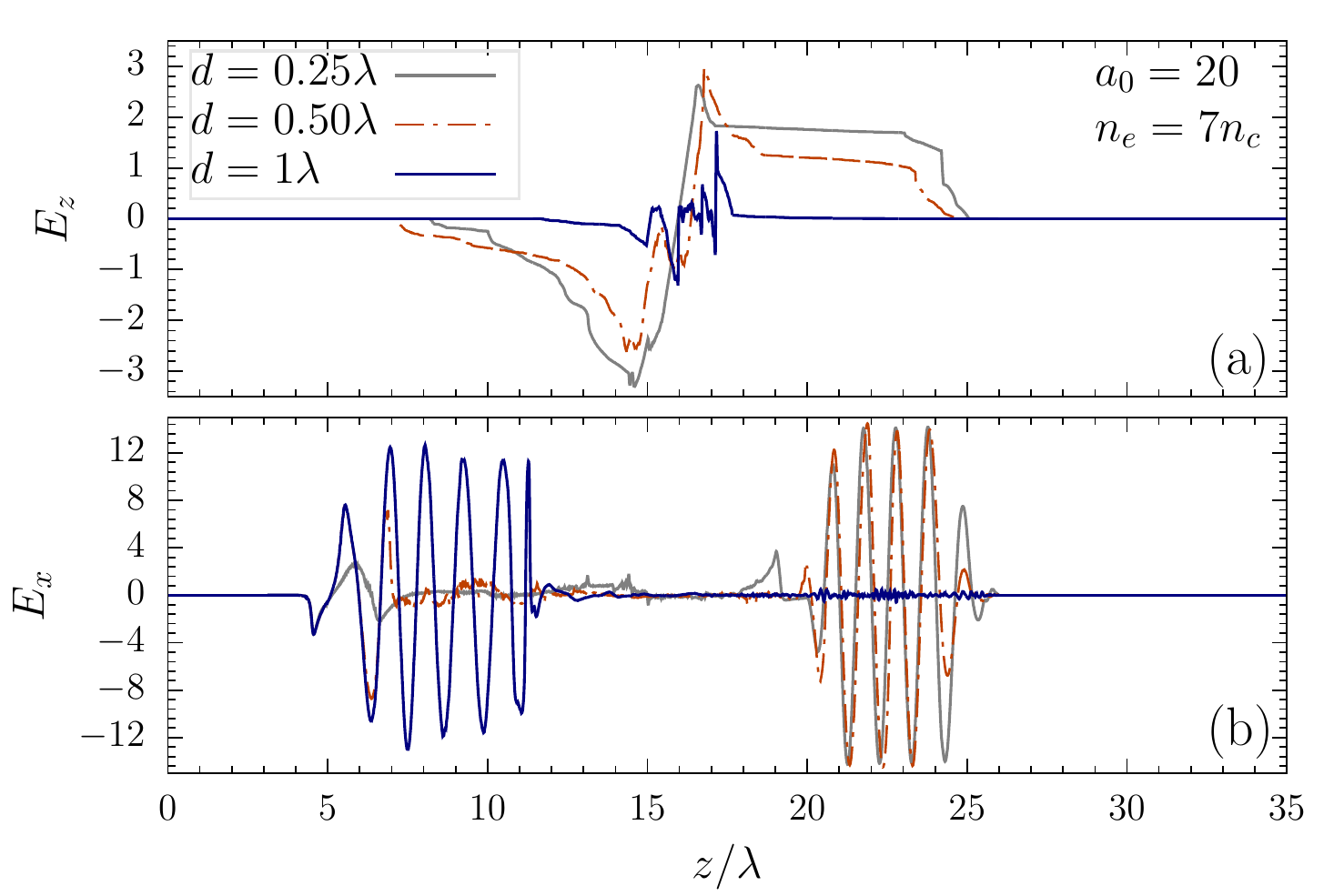}}
\caption{(color online) The spatial profile of electrostatic field ($E_z$) and laser field ($E_x$)  is plotted at 26$\tau$ for three different target thickness having density 7$n_c$. }
\label{ex_ez}
\end{figure}
  
\section{Results and Discussions}

In this paper we have used the 1D3V fully relativistic PIC code LPIC++ \cite{lpic} to carry out the studies. The dimensionless electric fields are normalized as $a_0 = eE/m_e\omega c$, where $\omega$ is the frequency of the laser pulse,   and $E$ is electric field amplitude in SI units. Time and space are normalized by one laser cycle and wavelength respectively. We have modified the code to be able to study the laser interaction with multilayer target structure. The laser pulse is considered to be a flat top of duration 6 cycles with rise and fall of 1 cycle each, propagating along the $z$ direction and incidents normally on the target. Throughout the paper we will be using a simulation domain of length $60\lambda$ (unless otherwise stated) and the Hydrogen plasma is considered to be located in region $15\lambda \leq z \leq d$, where $d$ is the target thickness.  Rest of the space is considered to be vacuum. The target density will be chosen such that RSIT will enable the laser to pass through. An extra layer of thickness 0.2$\lambda$ and having a density of $0.1n_c$ is placed behind the main target. The density of second layer is kept low so that the laser can also pass through this layer as well, and the ions from this layer will then see the electrostatic field created by RSIT and will accelerate as a bunch. We have used the term `ions' and `protons' interchangeably because we are only dealing with Hydrogen plasma in this paper.  

Before we start to study the effect of target thickness and density on RSIT and hence the accelerated ions, first we will see the difference between RSIT and LS regime of ion acceleration.

\begin{figure}[t]
\centerline{\includegraphics[scale=0.23]{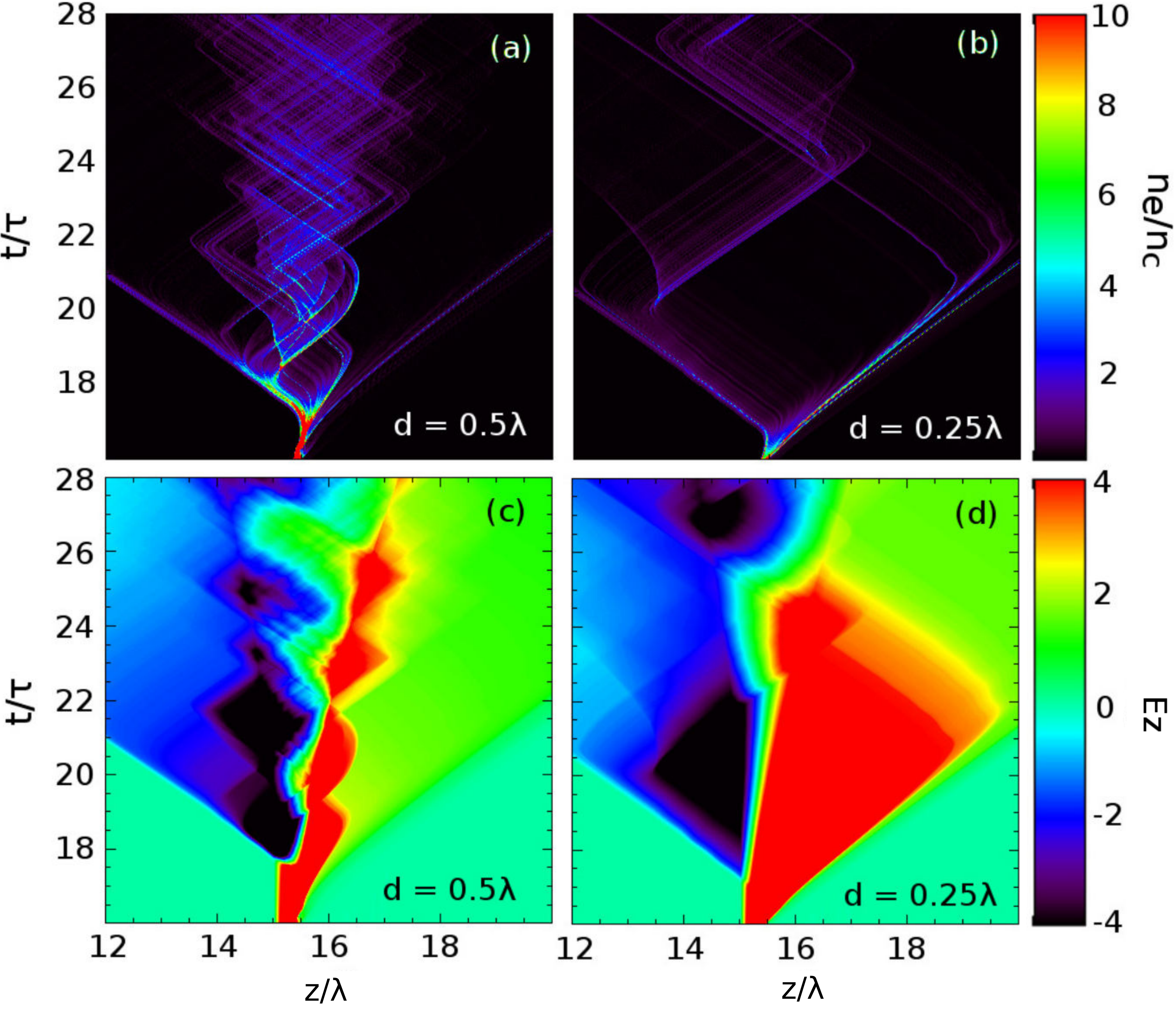}}
\caption{(color online) Space-time dependence of electron density for target thickness $0.5\lambda$ (a) and $0.25\lambda$ (b) along with longitudinal electrostatic field ($E_z$) for these two cases respectively are presented in (c) and (d).}
\label{thickness}
\end{figure}

\subsection{How RSIT is different from LS regime of acceleration ?}

As per as LS regime of ion acceleration is concerned, it has been reported that the thickness of the 
target foil plays quite an essential role in determining the energetics of accelerated ions \cite{macchi_ls,macchi_njp}. If the ratio $a_0/\xi < 1$, then the acceleration mechanism will be in LS regime and on the other hand the RSIT mechanism starts to prevail in the regime where the ratio $a_0/\xi \gtrsim 1$. Here, $a_0$ is the laser amplitude of circularly polarized laser and the parameter $\xi$ is defined as $\xi = \pi (n_e/n_c) (d/\lambda)$ with $n_e$, $n_c$ are respectively the target density and critical density corresponding to laser wavelength $\lambda$  and $d$ is the target thickness \cite{macchi_ls}.

\begin{figure*}[t]
\centerline{\includegraphics[scale=0.85]{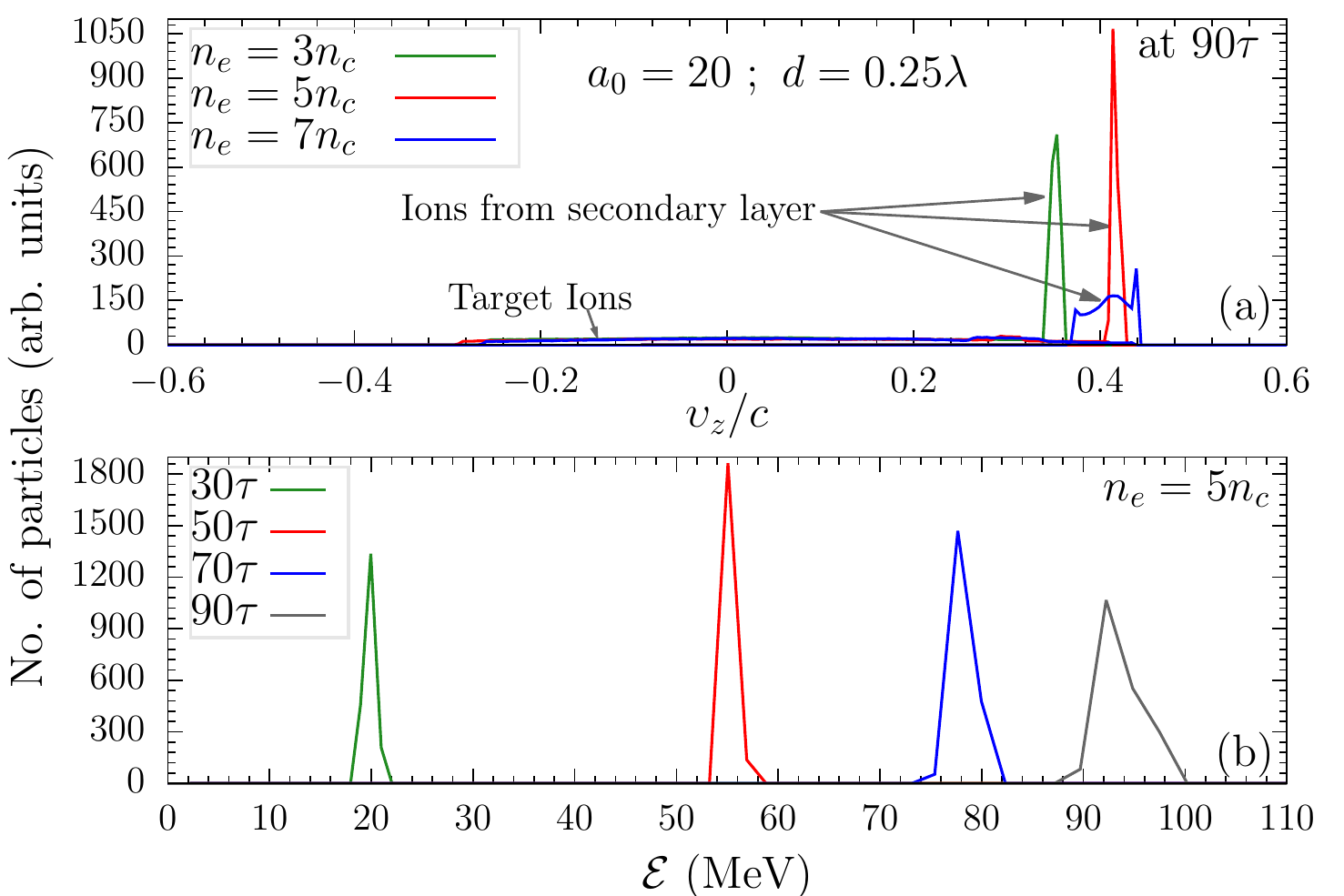}}
\caption{(color online) Velocity spectrum (a) of target ions and ions from secondary layer is measured at $90\tau$ for the case $a_0 = 20$, $d = 0.25\lambda$, $n_e = 3n_c$, $5n_c$ and $7n_c$. The energy spectrum (b) for $n_e = 5n_c$ at different time steps is also presented.}
\label{velspec}
\end{figure*}

\subsection{Formation of Electrostatic Field}
 
In the previous section we have discussed the basic theory behind the RSIT and elucidated the fact that the balance between the ponderomotive force by the laser and the electrostatic force by the charge separation is key and minimum criteria for RSIT mechanism. This balance of the forces plays paramount role especially when we have the plasma of the finite length. The electrostatic force originated by the charge separation is very much depend on the thickness as well as on density of the target. As an example, in Fig. \ref{threshold} we have presented the threshold density for RSIT as obtained by the PIC simulation for three different target thicknesses. It is being observed that the thin target supports comparatively larger target density as compare to the thicker target as per as RSIT is concerned. 

In order to see the effect of the target thickness on the formation of the electrostatic field, we have simulated the 
laser interaction with a target of thicknesses 0.25$\lambda$, 0.5$\lambda$ and 1$\lambda$. The laser conditions are 
same as in Fig. \ref{threshold} (circularly polarized, 6 cycle flat top with rise and fall of 1 cycle) with $a_0 = 20$
and target density for all the three cases is considered to be 7$n_c$. Figure \ref{ex_ez} shows the spatial profile of the electrostatic field (a) and laser field (b) at 26$\tau$. Here, $\tau$ is one laser cycle, and in vacuum laser propagates a distance of 1$\lambda$ in time 1$\tau$. As can be seen from Fig. \ref{ex_ez}(a) that for the case of 0.25$\lambda$ the 
electrostatic field is almost flat in the region just behind the target ($15\lambda\leq z \leq d$), while for the 
target with thickness 1$\lambda$, the strength of the electrostatic field is not that prominent. This can be 
explained on the basis of the threshold density for different target thickness as shown in Fig. \ref{threshold}. 
It can be read from Fig. \ref{threshold} that for a target which is 1$\lambda$ thick, the threshold density for $a_0 = 20$
lies slightly below the density used in this case (Fig. \ref{ex_ez}), however for 0.25$\lambda$ and 0.5$\lambda$, it is 
higher than 7$n_c$. This bring the target with thickness 0.25$\lambda$ and 0.5$\lambda$ in RSIT regime, which in principle is responsible for sweeping all the electrons and creating strong electrostatic field. The transparency of the target 
can also be seen from Fig. \ref{ex_ez}(b), where the reflection of laser pulse for 1$\lambda$ thickness and transmission for 0.25$\lambda$ and 0.5$\lambda$ thick target is clearly seen. It should be also noted that the parameter $\xi$ as defined earlier is about 5.5 and 11 for respectively $0.25\lambda$ and $0.5\lambda$ target thicknesses, and hence the interaction with $a_0 = 20$ brings it in RSIT regime, however, 1$\lambda$ thick target ($\xi \sim 22$) is opaque for the laser with $a_0 = 20$.
 
In Fig. \ref{ex_ez}, we have only shown the spatial profile of $E_z$ and $E_x$ only at 26$\tau$, however in order to see the spatio-temporal evolution of the electrostatic field, the electron density and corresponding electrostatic field for the case with $d = 0.25\lambda$ and $d = 0.5\lambda$ are presented in Fig. \ref{thickness}. As we are using the same density ($7n_c$) for thicknesses $0.25\lambda$ and $0.5\lambda$, hence its expected that the electrostatic force would be weaker (in comparison with ponderomotive force) in case of $0.25\lambda$ as compare to $0.5\lambda$, because effectively the target with thickness  $0.25\lambda$ would be having less charge particles as compare to the case when thickness is  $0.5\lambda$. The weaker electrostatic force in case of  $0.25\lambda$ manifests the stronger push on the electrons by CP laser pulse as can be seen in Fig. \ref{thickness}(b). On the other hand in case of $0.5\lambda$ the ponderomotive force of the laser is not strong enough to sweep the electrons to large distances as can be seen from Fig. \ref{thickness}(a). However, in both the cases the electrons are pushed away from the target, lowering the density of the target and hence enabling the target to be transparent for laser to pass through. When the laser is passed through the target then the electrons are pulled toward the target because of the electrostatic force created by charge separation of electrons and ions. As can be seen from Fig. \ref{thickness}(a), the electrons oscillate at higher frequency as compare to their low thickness counterpart as in Fig. \ref{thickness}(b). This can be understood from the lower excursion of the electrons in 0.5$\lambda$ case as compare to that of 0.25$\lambda$ case. It can be summarized as follows, for fixed density, thin (thick) target, will be having less (more) electrons, which results in weak (strong) electrostatic force to counter, and hence large (small) excursions of the electrons by ponderomotive force will take place, which in principle results in low (high) frequency electron oscillations by electrostatic force created by charge separation. The low frequency oscillations of the electron in case of thin target manifests the 
kind of constant electrostatic field which can be harnessed to accelerate the ions to very high energies.

\subsection{Acceleration}
 
So far we have discussed the formation of the electrostatic field by RSIT mechanism. Now we can exploit these fields to accelerate the protons to very high energies. Though it is clear that once the electrons are swept out from the target
then because of the space-charge effect of the ions from the target will start moving in either direction, which would
not yield in mono-energetic proton beam. In order to achieve the efficient acceleration by the electrostatic field created by RSIT, we have introduced a secondary layer of thickness 0.2$\lambda$ with density 0.1$n_c$ just behind the main target (laser incidents on the main target first). The density of this extra layer is considered to be low enough such that electrostatic fields created by RSIT mechanism is not affected by the presence of the secondary layer. Secondly, as a consequence of low target density, the laser will not have any direct effect on the energetics of the protons in this layer. 

\begin{figure}[b]
\centerline{\includegraphics[scale=0.63]{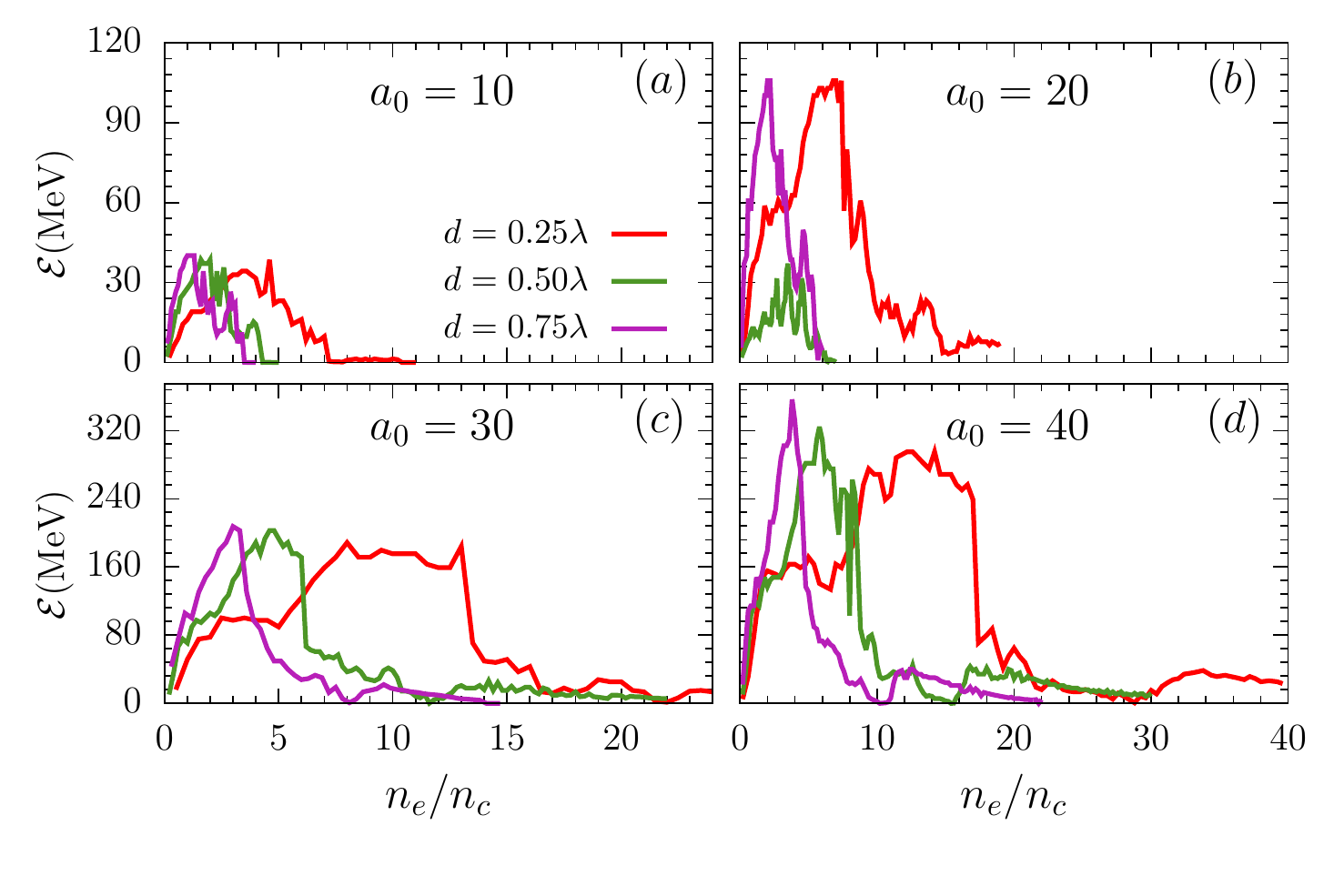}}
\caption{(color online) Variation of maximum energy of accelerated ions (from secondary layer) with density and thickness of primary layer for different laser amplitudes is presented.  }
\label{maxener}
\end{figure}

The velocity spectrum of the ions from the main target and the ions from secondary layer for the case $a_0 = 20$, $d = 0.25\lambda$, and $n_e = 3n_c$, $5n_c$ and $7n_c$ as evaluated at $90\tau$ are presented in Fig. \ref{velspec}(a). As can be seen from this figure that ions from secondary layer are getting accelerated as a bunch, on the other hand the ions from main target expands in either direction because of the space charge effect, resulting in large energy spread. This clearly shows the advantage of using an secondary layer behind the target to have a mono-energetic ion bunch. In this case the target ions are only responsible for creating an electrostatic field which is exploited by an additional thin layer of low density resulting in mono-energetic ion bunch. As per as the quality of the bunch is concerned, it can be seen that the spectra for the case $5n_c$ is much sharper than the one for the cases $3n_c$ and $7n_c$, which in fact signals toward the fact that there should be an optimum target density for a particular target thickness in order to have very efficient proton acceleration. The details regarding the optimum conditions for a given laser and target conditions is discussed in later part of this section.  
 
Furthermore, the energy spectrum of the accelerated protons from the secondary layer for the case  $n_e = 5n_c$ as evaluated at different times are presented in Fig. \ref{velspec}(b). As can be seen that at early stage of acceleration, the protons gain energy at faster rate and later the rate of acceleration decreases, this is mainly because of the weakening of the electrostatic force over a period of time. It is being observed that the ions from secondary layer are being accelerated to $\sim$ 100 MeV.  
 
\subsection{Optimization}

As it has been pointed out in Fig. \ref{velspec}  that depending on laser and target conditions, there should be some optimum conditions in order to accelerate the ions most efficiently. In view of this we have presented the variation of maximum kinetic energy of the secondary ions with  density and thickness of primary layer for different laser amplitudes in Fig. \ref{maxener}. The maximum value of the energy is evaluated at $90\tau$. It can be observed that there is an optimum density at which the ions can be accelerated most efficiently. This can be explained on the basis of the fact that for low densities overall charges itself is low enough resulting in weaker electrostatic field by primary target. On the other hand, for higher densities the RSIT mechanism ceases to exist because of reflection of the laser pulse and hence resulting in poor longitudinal electrostatic behind the target. In view of this it can be understood that there should be some optimum target density for particular target thickness at which the interplay between laser ponderomotive force, and the electrostatic forces by charge separation, results an ambient condition for the ions from secondary layer to be accelerated. The optimization of the target thickness for RSIT is also being reported in an experimental paper by Henig et. al. \cite{henig}.

\begin{figure}[t]
\centerline{\includegraphics[scale=0.6]{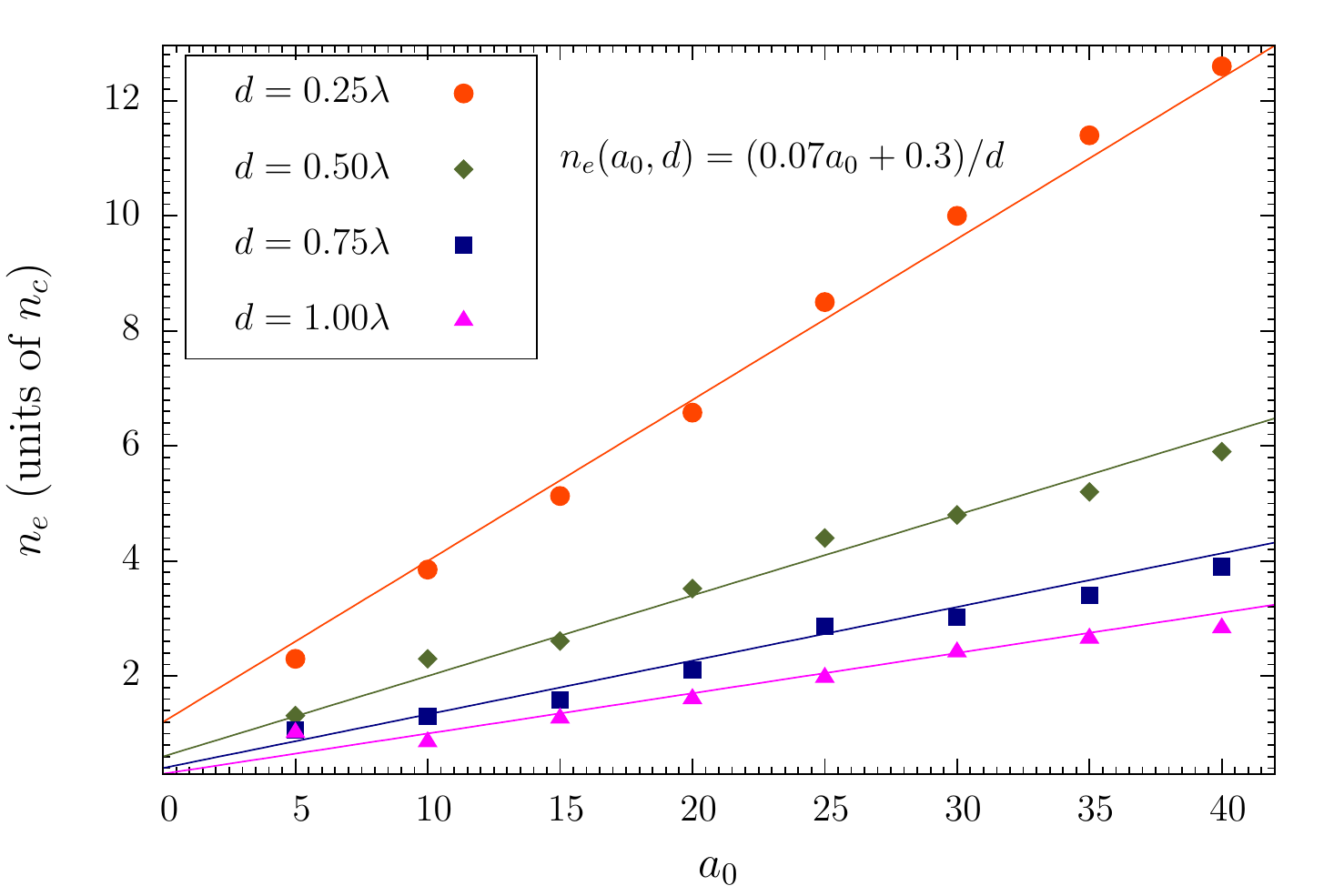}}
\caption{(color online) Optimum density of primary target is presented as a function of laser amplitude and its thickness. The solid line represents the scaling law for the optimum density as a function of laser amplitude and target thickness (Eq. \ref{scaling}). At these optimum densities the ions from secondary layer attains maximum energy.}
\label{optimum}
\end{figure}

The optimum density corresponding to particular laser and target conditions can be obtained from Fig. \ref{maxener}, e.g. for $a_0 = 20$, $d = 0.25\lambda$ the optimum density is about $\sim 6n_c$. In Fig. \ref{optimum} we have presented the optimum density as calculated by PIC simulations. In this figure we have presented the optimum density corresponding to particular thickness and density of primary layer, and laser amplitude. It has been observed that the optimum density for particular laser amplitude ($a_0$) and target thickness ($d$) can be fitted with following scaling law,
\be n_e(a_0,d) = (0.07 a_0 + 0.3) / d \label{scaling}\ee

As can be seen that Eq. (\ref{scaling}) fits very well with the predictions of the PIC simulations. The linear relation again hints towards the balance between the ponderomotive force ($\propto a_0$) and the electrostatic force ($\propto n_e d$) for maximum energy which in principle depends on the electrostatic force created by the charge separation in primary layer. In this 
work we have kept the thickness and density of secondary layer as 0.2$\lambda$ and $0.1n_c$ respectively, in principle one should also 
have some optimum conditions on secondary layer, but we would like to reserve it for the future.

\section{Concluding Remarks}
We have studied the RSIT from theoretical perspective by numerically solving the closed set of equations describing
the plasma response to intense laser fields, and results are compared against the 1D PIC simulations. The theoretical prescription tends to very well predict the strength of the longitudinal electrostatic field created by charge separation from laser ponderomotive force in semi-infinite plasma. The existing theoretical framework need to incorporate
the plasma expansion in vacuum in order to study the interaction of intense laser with target having finite thickness. In
order to study the RSIT in finite thickness target, we relied on the stationary plasma model to predict the threshold target
density for a particular laser amplitude and target thickness and found that for subwavelength targets the threshold density for RSIT is increased.
 
We have also discussed the role of target density and target thickness on accelerating fields and it is being observed that the protons from the secondary layer behind the target are accelerated most efficiently to few hundreds of MeV by electrostatic fields formed by RSIT. The role of target ions in this process is to create the electrostatic fields by RSIT mechanism, which will eventually accelerate the protons from the thin and low density layer placed behind the target. A suitable scaling law connecting the optimum target density for maximum energy, target thickness and laser amplitude is also being deduced. 

Though the results presented in this paper shed some light on the physics aspects of the acceleration of ions from secondary layer in RSIT enabled accelerating fields in subwavelength target. However, the detailed quantitative analysis and optimization of the physical parameters is still warranted, we have reserved that for our future work. We also would like to extend the presented theoretical framework for the targets with finite thickness. It should be also noted that we have used 1D PIC simulation to carry out these studies, and hence in order to experimentally realize these ideas one need to carry out full 3D PIC simulation of the same. 

\section*{Acknowledgements}
Authors acknowledge the Science and Engineering Research Board, Department of Science and Technology, Government of India for funding the project SR/FTP/PS-189/2012. Authors are immensely grateful to Gert Brodin for his valuable input.  

\appendix

\section{}
\label{App-A}
\begin{enumerate}
\item In coulomb gauge the Maxwell's equations will take the following form for vector potential $\mb{A}$ and 
scalar potential $\phi$,
\be \nabla^2 \phi = -\rho/\varepsilon_0\ee
\be \label{wave1} \nabla^2\mb{A} - \frac{1}{c^2} \ptdd{\mb{A}} = -\mu_0 \mb{J} + \frac{1}{c^2} \nabla \Big(\ptd{\phi}\Big) \ee
\be \mb{J} = -e n_e \mb{v_e} + Z e n_i \mb{v_i}\ee 
\be \mb{v_e} = v_z^e\ \hat{z} + \mb{v^e_\perp}\ee
\be \label{vi}\mb{v_i} = v_z^i\ \hat{z} + \mb{v^i_\perp}\ee
where, all the symbols have their usual meaning. The index $e$ and $i$ denote the corresponding quantity associated with 
electrons and ions respectively. The longitudinal motion is considered to be along the laser propagation direction ($z$), however transverse motion is associated along the direction of laser polarization.

\item Using 1D approximations the above equations can be written as ($z$ is direction of laser propagation), 
\be \psdd{\phi} =  -\rho/\varepsilon_0\ee
from Eq. \ref{wave1} to \ref{vi} after separating perpendicular and longitudinal component we can write [laser
pulse corresponds to vector potential $\mb{A}$ (transverse) and electrostatic field will correspond to $\phi$ (longitudinal)], 
\be \psdd{\mb{A}} - \frac{1}{c^2} \ptdd{\mb{A}} = \mu_0 (e n_e \mb{v^e_\perp} - Z e n_i \mb{v^i_\perp})\ee
\be \frac{1}{c^2} \pstd{\phi} =  -\mu_0 (e n_e  v^e_z - Z e n_i v^i_z)\ee

\item Now Consider the Lorentz force equations,
\be \ttd{\mb{P}} = q ( \mb{E} + \mb{v} \times \mb{B})\ee
\be \ttd{\mb{P}} = q \Big[-\nabla\phi - \ptd{\mb{A}} + \mb{v} \times (\nabla\times{\mb{A}})\Big]\ee
\be \ttd{\mb{P}} = q \Big[-\psd{\phi} - \ptd{\mb{A}} - v_z \psd{\mb{A}} + \Big(\mb{v_\perp} \cdot \psd{\mb{A}}\Big)\hat{z} \Big]\ee
again comparing longitudinal and transverse components,
\be \ttd{\mb{P_\perp}} = q \Big( -\ptd{\mb{A}} - v_z\psd{\mb{A}}\Big) = -q \ttd{\mb{A}} \implies \mb{P_\perp} = -q \mb{A} \ee 
\be  \mb{v_\perp} = \frac{-q\mb{A}}{m \gamma}\ ;\ \gamma = \sqrt{1 + \frac{\mb{P}^2}{m^2c^2}}\ee
\be \ttd{P_z} = -q \psd{\phi} + q \Big( \mb{v_\perp} \cdot \psd{\mb{A}}\Big)\ee
\be \ttd{P_z} = -q \psd{\phi} - \frac{q^2}{2 m \gamma} \psd{A^2}\ee
\be \ptd{P_z} + \frac{P_z}{m\gamma} \psd{P_z} = -q \psd{\phi} - \frac{q^2}{2 m \gamma} \psd{A^2}\ee

for electrons $q = -e$ and for ions $q = Z e$.

\item We will be using following dimensionless units,
\begin{enumerate}
 \item $\mb{a} = e\mb{A}/(m_e c)\ ; \  \varphi = e\phi/(m_e c^2) $
 \item $n_{e,i} = n_{e,i}/n_c\ ; \ n_c = \varepsilon_0 \omega^2 m_e/c^2 $  (critical density)
 \item $\mb{p} = \mb{P}/(m_e c)\ ; \ x = k x\ ; \ t = \omega t\ ; \ v = v / c\ ; \ m = m/m_e; \ q = q/e  $
  
\end{enumerate}
  
\item The equations in dimensionless units can be written as,
 \be \psdd{\varphi} =  n_e - Z n_i\ ; \psd{\varphi} = -E_z \ee
 \be \psdd{\mb{a}} -  \ptdd{\mb{a}} = \frac{n_e \mb{p_\perp^e}}{\gamma_e} - \frac{Z n_i \mb{p_\perp^i}}{\gamma_i m_i}\ee
 \be  \pstd{\varphi} =  -\ptd{E_z} = - \Big( n_e \frac{p_z^e}{\gamma_e} - Z n_i \frac{p_z^i}{\gamma_i m_i} \Big)\ee
 \be \ptd{E_z} =  \Big( n_e \frac{p_z^e}{\gamma_e} - Z n_i \frac{p_z^i}{\gamma_i m_i} \Big)\ee
 
 \be \mb{p_\perp^e} = \mb{a} \ ; \ \mb{p_\perp^i} = -Z \mb{a} \ee
 \be \psdd{\mb{a}} -  \ptdd{\mb{a}} = \Big(\frac{n_e}{\gamma_e} + \frac{Z^2 n_i}{\gamma_i m_i}\Big)\mb{a}\ee
 \be \label{gama}\gamma_e^2 = 1 + \mb{a}^2 + (p_z^e)^2 \ ;\ \gamma_i^2 = 1 + \frac{Z^2}{m_i^2}\mb{a}^2 + \Big(\frac{p_z^i}{m_i}\Big)^2 \ee
 \be \ptd{p_z^e} = \psd{\varphi} - \Big(\frac{1}{2\gamma_e} \psd{\mb{a}^2} + \frac{p_z^e}{\gamma_e} \psd{p_z^e}\Big) \ee  
 \be \ptd{p_z^i} = -Z \psd{\varphi} - \Big(\frac{Z^2}{2\gamma_i m_i} \psd{\mb{a}^2} + \frac{p_z^i}{\gamma_i m_i} \psd{p_z^i}\Big) \ee
 
 From Eq. \ref{gama} it can be proved that,
 \be \psd{\gamma_e} = \frac{1}{2\gamma_e} \psd{\mb{a}^2} + \frac{p_z^e}{\gamma_e} \psd{p_z^e}\ee
 \be \psd{\gamma_i} = \frac{Z^2}{2\gamma_i m_i^2} \psd{\mb{a}^2} + \frac{p_z^i}{\gamma_i m_i^2} \psd{p_z^i}\ee
 equations for $p_z$ will then be written as,
 \be \ptd{p_z^e} = -E_z - \psd{\gamma_e} \ ;\  \ptd{p_z^i} = Z E_z - m_i \psd{\gamma_i} \ee
 The above set of equations can be closed with continuity equations,
 \be \ptd{n_e} + \frac{\partial}{\partial z}\Big(n_e \frac{p_z^e}{\gamma_e}\Big) = 0\ ; \ 
 \ptd{n_i} + \frac{\partial}{\partial z}\Big(n_i \frac{p_z^i}{\gamma_i m_i}\Big) = 0 \ee

\item Now final closed set of equations to be solved are,
  
  \be \psd{E_z} = Z n_i - n_e \ee
  \be \ptd{E_z} =  \Big( n_e \frac{p_z^e}{\gamma_e} - Z n_i \frac{p_z^i}{\gamma_i m_i} \Big)\ee
  \be \psdd{\mb{a}} -  \ptdd{\mb{a}} = \Big(\frac{n_e}{\gamma_e} + \frac{Z^2 n_i}{\gamma_i m_i}\Big)\mb{a}\ee
  \be \ptd{p_z^e} = -E_z - \psd{\gamma_e} \ee
  \be \ptd{p_z^i} = Z E_z - m_i \psd{\gamma_i} \ee
  \be \ptd{n_e} + \frac{\partial}{\partial z}\Big(n_e \frac{p_z^e}{\gamma_e}\Big) = 0 \ee
  \be \ptd{n_i} + \frac{\partial}{\partial z}\Big(n_i \frac{p_z^i}{\gamma_i m_i}\Big) = 0 \ee
  \be \gamma_e  = \sqrt{1 + \mb{a}^2 + (p_z^e)^2} \ee
  \be \gamma_i = \sqrt{1 + \frac{Z^2}{m_i^2}\mb{a}^2 + \Big(\frac{p_z^i}{m_i}\Big)^2} \ee
 
\item If we consider the steady ionic background, then the closed set of equations reduces to,
  \be \psd{E_z} = Z n_i - n_e \ee
  \be \ptd{E_z} = n_e \frac{p_z^e}{\gamma_e}  \ee
  \be \psdd{\mb{a}} -  \ptdd{\mb{a}} =  \frac{n_e}{\gamma_e} \mb{a}\ee
  \be \ptd{p_z^e} = - E_z - \psd{\gamma_e} \ee
  \be \ptd{n_e} + \frac{\partial}{\partial z}\Big(n_e \frac{p_z^e}{\gamma_e}\Big) = 0 \ee
  \be \gamma_e  = \sqrt{1 + \mb{a}^2 + (p_z^e)^2} \ee

 \end{enumerate}

\end{document}